\documentclass[a4paper]{article}

\usepackage{graphicx}
\begin{document}

\title{Scheme for generating entangled states of two field modes in a cavity}
\author{J. LARSON \\ Physics
  Department\\ Royal Institute of Technology (KTH)\\ Albanova,
  Roslagstullsbacken 21\\ SE-10691 Stockholm, Sweden}

\maketitle

\begin{abstract}
This paper considers a two-level atom interacting with two cavity
modes with equal frequencies. Applying a unitary transformation,
the system reduces to the analytically solvable Jaynes-Cummings
model. For some particular field states, coherent and squeezed
states, the transformation between the two bare basis's, related
by the unitary transformation, becomes particularly simple. It is
shown how to generate, the highly non-classical, entangled
coherent states of the two modes, both in the zero and large
detuning cases. An advantage with the zero detuning case is that
the preparation is deterministic and no atomic measurement is
needed. For the large detuning situation a measurement is
required, leaving the field in either of two orthogonal entangled
coherent states.
\end{abstract}

\section{Introduction}\label{intro}
Quantum information processing and nonlocality tests rely on the
phenomenon of entenglement \cite{qinf}. This is a purely quantum
mechanical correlation between different subsystems, for example,
atoms or photons. During the last decades, entanglement has not
only been an issue for theoreticians but entangled states have
been prepared and measured in various experimental set-ups. Many
of the successful demonstrations have been within quantum optical
systems \cite{leshouches}. One of the prominent subfields is
cavity quantum electrodynamics (cavity QED). In cavity QED, single
atomic transitions and single cavity modes can be isolated and
coupled coherently, and, for example, preparation of entangled
states is achievable \cite{ent}. The atom constitute a two-level
system, while the cavity mode a harmonic oscillator, and due to
the large atom-field coupling together with long lifetimes of the
atomic Rydberg states and the use of a high $Q$ cavity,
decoherence effects can often be neglected during the interaction
time \cite{cqed}. Usually the atom is sent through the cavity and
interacts resonantly or nearly resonantly with one of the cavity
modes. The dynamics is, in most cases, well described by the
analytically solvable Jaynes-Cummings (JC) model \cite{jc}.

Schr\"odinger cat states \cite{cat}, the superposition of two
coherent states with large amplitudes, have achieved great
theoretical and experimental attention, since they are believed to
approach semi-classical regions. The properties of such states
have been studied in great detail, see {\it e.g.} \cite{catprop}.
Entangled coherent states \cite{multicat} are the generalization
of cat states for several modes. These poses the non-classical
feature, apart from the superposition, of entanglement. Especially
the states of two modes have been investigated \cite{2multicat}
and their properties, for example, entanglement
\cite{multicat,multicatent}, decoherence \cite{multicatdec,qcom}
(see also \cite{2moddec}) and non-classical effects
\cite{multicatprop}. One of the reason for the extensive interest
of these states are for their applicability within quantum
information processing, see the review article \cite{qcom} and
references therein. For example, quantum teleportation
\cite{multicatdec,cohtel}, quantum computing \cite{cohcomp},
non-locality tests \cite{nonlocal} and quantum communication
\cite{cohcommun}. Suggested preparation schemes of entangled
coherent states have been covered in a large amount of articles
\cite{qcom}. The proposed methods can be divided into: free
propagating waves and linear optical elements \cite{lin}, Kerr
non-linearities \cite{kerr}, trapped ions \cite{ion} and cavity
QED \cite{cavity1,cavity2,cavity3,jonasbarry,erikajonas}.

The intent of this paper is to present a method, achievable in
todays experiments \cite{gb2,haroche1,haroche2}, for the
generation of entangled coherent states within cavity QED. The
model I study here is an extended JC model with a two-level atom
interacting with two cavity modes \cite{2mod1,fredrik,
marcelo,plk0}. In the case of modes with equal frequencies, the
model can be transformed into the standard JC model by a unitary
operation. Thus the system is easily  solvable in the new
transformed basis. The relation between the standard bare state
basis and the new quasi bare state basis is not trivial for any
fields. However, we show that for coherent and squeezed state
bases a simple transformation relates the two, which are key
results of the paper. It is analytically shown how entangled
coherent states can be generated in the zero and large atom-field
detuning cases. This is done by the knowledge of how the standard
JC model evolves for coherent field states \cite{leshouches,gb}.
One advantage with the zero detuning scheme is that no atomic
measurement is needed, while in the large detuning case a
measurement is demanded, which, however, always leaves the field
in either of two orthogonal entangled coherent states. Two known
approximations are used in deriving the dynamics; the large
amplitude approximation, and adiabatic elimination of one of the
atomic levels. The former has, to my knowledge, not been
considered for the model used in this paper. The latter, adiabatic
elimination approximation, is considered in \cite{cavity1}, also
for the preparation of entangled coherent states. The effective
Hamiltonian used in \cite{cavity1} is not the one obtained if the
elimination is carried out correctly. The 'false' Hamiltonian
does, however, give similar results, and it is explained why this
comes about. I give the correct Hamiltonian and discuss how this
may generate entangled coherent states, also in the case of
different mode frequencies. In \cite{marcelo}, analytical
expressions are derived for the atomic inversion and phase space
distributions, but as a state preparation process it is not
discussed. Further, in \cite{plk0}, the same model is used for
homodyne detection of the cavity field.

The outline of the paper is: In section \ref{sec2} the model is
introduced and the unitary transformation giving the standard JC
model, and then I derive the relations between the original and
the quasi modes for coherent and squeezed states. Section
\ref{sec3} considers the zero detuning situation and briefly
reviews the dynamics of the standard JC model in the presence of
coherent states with large amplitudes. I then show how, within
these approximations, the evolution of the combined system may
lead to entangled coherent states of the two modes. The following
section \ref{sec4} uses the effective Hamiltonian obtained by
adiabatic elimination of one of the two atomic states due to a
large atom-field detuning. This Hamiltonian governs an evolution
that also results in entangled coherent states. Finally I conclude
the paper in section \ref{conc} with some comments.

\section{Describing the model}\label{sec2}
The JC model \cite{jc} has served as a theoretical description of
the interaction between a single atom and a single mode inside a
high $Q$ cavity. In these situations only one transition of the
atom may be considered, reducing the atom degrees of freedom to a
two-level system. The analytically solvable JC model defines a
two-level system (spin 1/2) coupled to a harmonic oscillator,
within the dipole and rotating wave approximation. In this paper I
consider the extended Jaynes-Cummings model, including one
two-level atom and two cavity modes, 1 and 2. The Hamiltonian,
after adding a second cavity mode, reads ($\hbar=1$)
\begin{equation}\label{ham0}
\tilde{H}=\!\frac{\Omega}{2}\sigma_z+\omega_1a^\dagger\!
a+\omega_2b^\dagger\!
b+\!\big(g_1a+g_2b\big)\sigma^+\!+\big(g_1^*a^\dagger\!+g_2^*b^\dagger\big)\sigma^-.
\end{equation}
Here the sigma-operators act on the atomic ground and excited
states $|-\rangle$ and $|+\rangle$: $\sigma_z=|+\rangle\langle
+|-|-\rangle\langle -|$, $\sigma^+=|+\rangle\langle -|$ and
$\sigma^-=|-\rangle\langle +|$. The two operators $a$
($a^\dagger$) and $b$ ($b^\dagger$) are regular boson operators
acting on mode 1 and 2 respectively:
$[a,a^\dagger]=1=[b,b^\dagger]$ and $[a,b]=[a^\dagger,b]=...=0$.
The two couplings are given by $g_{1,2}$, the mode frequencies by
$\omega_{1,2}$ and the atomic transition frequency by $\Omega$.
Bare states are given by $|n\rangle_1|m\rangle_2|\pm\rangle$,
where the first two kets give the number of photons in mode 1 and
2 respectively, $a^\dagger
a|n\rangle|m\rangle|\pm\rangle=n|n\rangle|m\rangle|\pm\rangle$ and
$b^\dagger
b|n\rangle|m\rangle|\pm\rangle=m|n\rangle|m\rangle|\pm\rangle$,
while the last one gives the atomic state,
$\sigma_z|n\rangle|m\rangle|\pm\rangle=\pm|n\rangle|m\rangle|\pm\rangle$.
The number of excitations are preserved by the
Hamiltonian~(\ref{ham0}); $[N,H]=0$, where
$N=\frac{1}{2}\sigma_z+a^\dagger a+b^\dagger b$. It is convenient
to work in the interaction picture with respect to the operator
$N$, and from now on we assume $\omega_1=\omega_2=\omega$ and the
couplings $g_{1,2}$ to be real, giving the interaction picture
Hamiltonian
\begin{equation}
H=\tilde{H}-\omega
N=\frac{\Delta}{2}\sigma_z+\big(g_1a+g_2b\big)\sigma^++\big(g_1a^\dagger+g_2b^\dagger\big)\sigma^-,
\end{equation}
 where $\Delta=\Omega-\omega$ is the atom-field detuning. In order to proceed I introduce the new boson operators $A$ and $B$, acting on quasi modes $I$ and $II$ respectively, by a orthogonal transformation $U$ according to
\begin{equation}\label{trans}
\left[\begin{array}{c} A \\
B\end{array}\right]=U\left[\begin{array}{c} a \\
b\end{array}\right],\hspace{0.75cm}U=\left[\begin{array}{cc}
\cos(\theta) & \sin(\theta) \\ -\sin(\theta) &
\cos(\theta)\end{array}\right],
\end{equation}
where $\cos(\theta)=g_1/g$ and $g=\sqrt{g_1^2+g_2^2}$. With the
boson operators $A$ and $B$, the Hamiltonian has the form of a
standard JC one for quasi mode $I$,
\begin{equation}
H=\frac{\Delta}{2}\sigma_z+g\big(A^\dagger\sigma^-+A\sigma^+\big).
\end{equation}
Note that $a^\dagger a+b^\dagger b=A^\dagger A+B^\dagger B$,
meaning that $\Delta$ is also the detuning between the atom and
quasi fields $I$ and $II$. The number, or Fock, states for the
quasi modes are written as $|N\rangle_I$ and $|M\rangle_{II}$, and
it follows that quasi mode $II$ is unaffected by the Hamiltonian
$H$ in the interaction picture. While in the non interaction
pictures it accumulate a phase proportional to $\omega B^\dagger
B$. In the zero detuning case, an initial  state
\begin{equation}\label{instate}
|\Psi(0)\rangle=\sum_{N,M}C_N^{(I)}C_M^{(II)}|N\rangle_I|M\rangle_{II}\big[\gamma|-\rangle+\delta|+\rangle\big]
\end{equation}
evolves into
\begin{equation}\label{timesol}
\begin{array}{lll}
|\Psi(t)\rangle & = &
\left[\sum_MC_M^{(II)}|M\rangle_{II}\right]\frac{1}{\sqrt{2}}\sum_{N}\left[\left(\gamma
C_{N+1}^{(I)}+\right.\right.\\ \\ & & \left.\left. +\delta
C_{N}^{(I)}\right)\mathrm{e}^{igt\sqrt{N+1}}|\phi_{1N}\rangle_I\right.
\\ \\ & & \left.+\left(\delta C_{N}^{(I)}-\gamma
C_{N+1}^{(I)}\right)\mathrm{e}^{-igt\sqrt{N+1}}|\phi_{2N}\rangle_I\right],
\end{array}
\end{equation}
where we have introduced the quasi dressed states
\begin{equation}
\begin{array}{c}
|\phi_{1N}\rangle_I=\frac{1}{\sqrt{2}}\big[|N\rangle_I|+\rangle+|N+1\rangle_I|-\rangle\big]\\ \\
|\phi_{2N}\rangle_I=\frac{1}{\sqrt{2}}\big[|N\rangle_I|+\rangle-|N+1\rangle_I|-\rangle\big].
\end{array}
\end{equation}
The particular case of single quasi Fock states,
$C_N^{(I)}=\delta_{NN'}$, has been studied in \cite{2mod1}, and
here I consider coherent or squeezed states. The displacement
operator $D(\alpha)=\exp\left(\alpha a^\dagger-\alpha^* a\right)$,
creates a coherent state with amplitude $\alpha$ when it operates
on vacuum, $D(\alpha)|0\rangle=|\alpha\rangle$. Likewise, the
operators $A$ and $B$ may be used to define displacement operators
for quasi modes $I$ and $II$. If both modes 1 and 2 are prepared
in coherent states, it is easy to find corresponding states of the
quasi modes $I$ and $II$ using eq (\ref{trans}),
\begin{equation}\label{relation1}
\begin{array}{lll}
D_1(\alpha)D_2(\beta) & = & \exp\left(\alpha
a^\dagger-\alpha^*a+\beta b^\dagger-\beta^*b\right) \\ \\ & = &
\exp\left[\alpha\left(\cos(\theta)A^\dagger-\sin(\theta)B^\dagger\right)\right.
\\ \\& & -\alpha^*\left(\cos(\theta)A-\sin(\theta)B\right)\\ \\ &
& +\beta\left(\sin(\theta)A^\dagger+\cos(\theta)B^\dagger\right)\\
\\ & &
-\left.\beta^*\left(\sin(\theta)A+\cos(\theta)B\right)\right]\\ \\
& = &
\exp\left[\left(\alpha\cos(\theta)+\beta\sin(\theta)\right)A^\dagger\right.
\\ \\ & & -\left(\alpha^*\cos(\theta)+\beta^*\sin(\theta)\right)A
\\ \\ & &
+\left(\beta\cos(\theta)-\alpha\sin(\theta)\right)B^\dagger\\ \\ &
&
-\left.\left(\beta^*\cos(\theta)-\alpha^*\sin(\theta)\right)B\right]\\
\\ & = & D_I(\mu)D_{II}(\nu),
\end{array}
\end{equation}
 where
\begin{equation}\label{amprel}
\left[\begin{array}{c}\mu \\
\nu\end{array}\right]=\left[\begin{array}{cc}\cos(\theta) &
\sin(\theta)  \\ -\sin(\theta)
& \cos(\theta)\end{array}\right]\left[\begin{array}{c}\alpha \\
\beta\end{array}\right].
\end{equation}
Thus, the amplitudes $\alpha$, $\beta$, $\mu$ and $\nu$ are
related in the same way as the boson operators $a$, $b$, $A$ and
$B$. It is understood that the relation (\ref{relation1}) works
both ways, $1,2\leftrightarrow I,II$. A special case of this
relation between the basis is mentioned in \cite{plk0}

Let us now derive the corresponding relation between the squeezing
operators
$S(z)=\exp\left[\frac{1}{2}\left(z^*a^2-za^{\dagger2}\right)\right]$,
see \cite{mandle}, for the various modes. Using the
Campell-Baker-Hausdorff theorem \cite{mandle},
$\exp(A+B)=\exp(A)\exp(B)\exp\left(-[A,B]/2\right)$, where the
operators $A$ and $B$ fulfills $[A,[A,B]]=0=[B,[A,B]]$, it follows
\begin{equation}\label{relation2}
S_1(z_1)S_2(z_2)=\mathrm{e}^{\left(p^*AB-pA^\dagger
B^\dagger\right)}S_I(q)S_{II}(q),
\end{equation}
where $p=\sin(2\theta)(z_2-z_1)/2$ and
$q=z_1\cos^2(\theta)+z_2\sin^2(\theta)$. When $z_1=z=z_2$, $p=0$
and $q=z$, and the first exponential on the right hand side
disappears, resulting in two ordinary squeezing operators for mode
$I$ and $II$.

The results presented in eqs. (\ref{relation1}), (\ref{amprel})
and (\ref{relation2}) are a main underlying condition for this
paper; an entangled coherent or Schr\"odinger cat state in one
basis will give a similar state in the other basis. This might be
an intuitive result, but one should remember that the unitary
transformation (\ref{trans}) for making the 2-mode JC model
(\ref{ham0}) into a standard JC model, has the simple form since
the two modes have identical frequencies. In section \ref{sec4}
for large detunings, it will be shown that the generation of
entangled coherent states for modes with different frequencies is
more complicated.

\section{Generation of entangled states 1; zero detuning}\label{sec3}

In the previous section it was shown that having two coherent
states in modes 1 and 2 is the same as having two coherent states
in modes $I$ and $II$ with different amplitudes. The dynamics of
the original JC model in the case of a coherent state with a large
amplitude is well understood \cite{gb2,gb}, see also
\cite{plk,plk2}. Surprisingly, in the large amplitude limit it was
found that, independently of the atoms initial state it
disentangles from the field after a particular time, the half
revival time $t_R=2\pi\sqrt{\bar{n}}/g$ where $\bar{n}$ is the
average photon number. At this time the initial coherent state has
split up into a superposition of two coherent states with opposite
phases, a so called Schr\"odinger cat state. I briefly review this
phenomenon here and for a more rigorous mathematical treatment see
\cite{gb}.

For initial coherent states we have
\begin{equation}
C_{N+1}=\mathrm{e}^{-i\varphi}\left(\frac{\bar{N}}{N+1}\right)^{1/2}C_N,
\end{equation}
where $\varphi$ is the phase of the field;
$\mu=\sqrt{\bar{N}}\mathrm{e}^{-i\varphi}$. In the limit of large
$\bar{N}$, it is legitimate to approximate
$C_{N+1}\approx\mathrm{e}^{-i\varphi}C_N$. Since, for the coherent
state with  large amplitude, the distribution $C_N$ is sharply
peaked around its average $\bar{N}$, I expand the exponent
accordingly
\begin{equation}
\begin{array}{c}
\sqrt{N+1}\approx \sqrt{N}+\frac{1}{2\sqrt{N}}\approx \sqrt{N}+\frac{1}{2\sqrt{\bar{N}}}, \\ \\
\sqrt{N}\approx\frac{\sqrt{\bar{N}}}{2}+\frac{N}{2\sqrt{\bar{N}}}-\frac{(N-\bar{N})^2}{8\bar{N}^{3/2}}...
\end{array}
\end{equation}
For simplicity, assume $\gamma=1$, then the time-dependent
solution (\ref{timesol}), using the above approximations, reduces
to
\begin{equation}\label{appsol}
\begin{array}{lll}
|\Psi(t)\rangle & \approx &
\left[\sum_MC_M^{(II)}|M\rangle_{II}\right]\frac{1}{2}\left[\mathrm{e}^{igt\sqrt{\bar{N}}/2}\sum_{N}C_N^{(I)}\times\right.
\\ \\ & & \mathrm{e}^{igtN/2\sqrt{\bar{N}}}|N\rangle_I\left(\mathrm{e}^{igt/2\sqrt{\bar{N}}}|+\rangle+\mathrm{e}^{i\varphi}|-\rangle\right) \\ \\
& & +
\mathrm{e}^{-igt\sqrt{\bar{N}}/2}\sum_{N}C_N^{(I)}\mathrm{e}^{-igtN/2\sqrt{\bar{N}}}|N\rangle_I\times\\
\\ & &
\left.\left(\mathrm{e}^{-igt/2\sqrt{\bar{N}}}|+\rangle-\mathrm{e}^{i\varphi}|-\rangle\right)\right]
\end{array}
\end{equation}
Now, at $t=t_R/2$ it is clear that the atomic and field states
separates, and for a coherent distribution in quasi mode $I$ with
amplitude $\mu$, the state becomes
\begin{equation}
\begin{array}{lll}
|\Psi(t\!=\!t_R/2)\rangle & \! =\! & \!
\left[\sum_M\!C_M^{(II)}\!|M\rangle_{II}\!\right]\!\frac{1}{\sqrt{2}}\!\left(i|+\rangle\!+\!\mathrm{e}^{i\varphi}|-\rangle\right)\!\times
\\ \\ & &
\frac{1}{\sqrt{2}}\left(\mathrm{e}^{i\pi\bar{N}}|i\mu\rangle_I-\mathrm{e}^{-i\pi\bar{N}}|-i\mu\rangle_I\right),
\end{array}
\end{equation}
where we assume $\langle-i\mu|i\mu\rangle\approx0$ in the large
amplitude limit. Thus, within the validity of the approximations,
as the atom disentangle from the field, the field has split up
into a superposition of coherent states with opposite phases. The
interesting fact is that the disentanglement appears for any
initial atomic state, the similar calculation as in eq.
(\ref{appsol}) could be performed for $\delta=1$ and since
$|\phi_{1,2}\rangle$ span the whole atomic subspace, it holds for
any initial state. By using squeezed states which may be more
peaked around their means $\bar{N}$, the validity constrains are
more easily fulfilled \cite{jonasbarry,plk}. The similar splitting
of the field state in phase space is present for a squeezed state.

Going back to our extended JC model with two modes, we assume that
modes 1 and 2 are initially in coherent states with large
amplitudes $\alpha$ and $\beta$. The quasi modes $I$ and $II$ will
then be in coherent states with amplitudes $\mu$ and $\nu$
according to eq. (\ref{amprel}). By letting the atom interact for
a time $t=t_R/2$, where $t_R$ is the revival time for the quasi
mode $I$, $t_R=2\pi\sqrt{\bar{N}}/g$, the field becomes
\begin{equation}
|\psi_{f}(t_R/2)\rangle=\frac{1}{\sqrt{2}}\left(\mathrm{e}^{i\pi\bar{N}}|i\mu\rangle_I-\mathrm{e}^{-i\pi\bar{N}}|-i\mu\rangle_I\right)|\nu\rangle_{II}.
\end{equation}
Going to modes 1 and 2 we get
\begin{equation}
|\psi_{f}(t_R/2)\rangle=\frac{1}{\sqrt{2}}\left(\mathrm{e}^{i\pi\bar{N}}|\alpha'\rangle_1|\beta'\rangle_2-\mathrm{e}^{-i\pi\bar{N}}|\alpha''\rangle_1|\beta''\rangle_2\right),
\end{equation}
where
\begin{equation}\label{amprel2}
\begin{array}{l}
\left[\!\begin{array}{c} \alpha' \\ \beta'\end{array}\!\right]\!\!=\!\!\left[\!\begin{array}{cc} \cos(\theta) & -\sin(\theta) \\
\sin(\theta) &Ê\cos(\theta)\end{array}\!\right]\!\left[\!\begin{array}{cc} i & 0 \\ 0 & 1\end{array}\!\right]\!\left[\!\begin{array}{cc} \cos(\theta) & \sin(\theta) \\
-\sin(\theta) &Ê\cos(\theta)\end{array}\!\right]\!\left[\!\begin{array}{c} \alpha \\ \beta\end{array}\!\right] \\ \\
\left[\!\begin{array}{c} \alpha'' \\ \beta''\end{array}\!\right]\!\!=\!\!\left[\!\begin{array}{cc} \cos(\theta) & \!-\sin(\theta) \\
\sin(\theta) &Ê\cos(\theta)\end{array}\!\!\right]\!\left[\!\!\begin{array}{cc} -i & 0 \\ 0 &Ê1\end{array}\!\right]\!\left[\!\!\begin{array}{cc} \cos(\theta) & \sin(\theta) \\
-\sin(\theta)
&Ê\cos(\theta)\end{array}\!\!\right]\!\left[\!\begin{array}{c} \alpha \\
\beta\end{array}\!\right]\!\!.
\end{array}
\end{equation}
For $\alpha=-i\sqrt{\bar{n}}=\beta$ and  $g_1=g_2$, the angle $\theta$ is $\pi/2$, giving
\begin{equation}
|\psi_{f}(t_R/2)\rangle\!=\!\frac{1}{\sqrt{2}}\!\left(\mathrm{e}^{i2\pi\bar{n}}|\bar{n}\rangle_1|\bar{n}\rangle_2-\mathrm{e}^{-i2\pi\bar{n}}|-\bar{n}\rangle_1|-\bar{n}\rangle_2\right)\!.
\end{equation}

\section{Generation of entangled states 2; large detuning}\label{sec4}
When the atom-field detuning is large compared to the coupling,
population is unlikely to transfer between the internal atomic
states $|\pm\rangle$. It is then possible to approximate the JC
model by an effective one where the initially empty level has been
adiabatically eliminated. In the appendix \ref{app} I derive an
effective JC model by transforming the Hamiltonian by an  unitary
operator \cite{adel1,adel2}. The resulting Hamiltonian, to order
$\mathcal{O}(g^2/\Delta)$, becomes
\begin{equation}\label{effH2}
H_{el}=\left[\frac{\Delta}{2}+\frac{g^2}{\Delta}A^\dagger A\right]\sigma_z+\frac{g^2}{2\Delta}\sigma^+\sigma^-.
\end{equation}
Evolving an initial state (\ref{instate}) by this Hamiltonian gives
\begin{equation}
\begin{array}{lll}
|\Psi(t)\rangle & = & U(t)|\Psi(0)\rangle=\left[\sum_{M}C_M^{(II)}|M\rangle_{II}\right]\times \\ \\
& &
\times\sum_N\left[C_N^{(I)}\gamma\mathrm{e}^{-i\left(\frac{\Delta}{2}+\frac{g^2N}{\Delta}\right)t}|N\rangle_I|-\rangle\right. \\
\\ & &
\left.+\delta\mathrm{e}^{+i\left(\frac{\Delta}{2}+\frac{g^2}{2\Delta}+\frac{g^2N}{\Delta}\right)t}|N\rangle_I|+\rangle\right].
\end{array}
\end{equation}
And for initial coherent states with amplitudes $\mu$ and $\nu$ we get
\begin{equation}
|\Psi(t)\rangle\!=\!\left[\!\gamma|\mu\mathrm{e}^{-i\frac{g^2t}{\Delta}}\rangle_I|-\rangle\!+\!\delta\mathrm{e}^{i\left(\Delta+\frac{g^2}{2\Delta}\right)t}|\mu\mathrm{e}^{i\frac{g^2t}{\Delta}}\rangle_I|+\rangle\!\right]\!|\nu\rangle_{I\!I}.
\end{equation}
Assume $\gamma=\delta=1/\sqrt{2}$ and that the atom is measured at
time $t'$ in the $\frac{1}{\sqrt{2}}(|-\rangle\pm|+\rangle)$
basis, leaving the fields in the states
\begin{equation}
\begin{array}{l}
|\psi_{f}(t')\rangle\!=\!\frac{1}{\sqrt{2}}\!\left[\!|\mu\mathrm{e}^{-i\frac{g^2t'}{\Delta}}\rangle_I\!|\nu\rangle_{I\!I}\!+\!\mathrm{e}^{i\left(\!\Delta+\frac{g^2}{2\Delta}\right)\!t'}\!|\mu\mathrm{e}^{i\frac{g^2t'}{\Delta}}\rangle_I\!|\nu\rangle_{I\!I}\!\right]\!\!, \\ \\
|\psi_{f}(t')\rangle=\frac{1}{\sqrt{2}}\left[|\alpha'\rangle_1|\beta'\rangle_{2}+\mathrm{e}^{i\left(\Delta+\frac{g^2}{2\Delta}\right)t'}|\alpha''\rangle_1|\beta''\rangle_{2}\right],
\end{array}
\end{equation}
where the amplitudes $\alpha'$, $\beta'$, $\alpha''$ and $\beta''$
are given by eq. (\ref{amprel2}) with the middle matrix changed to
\begin{equation}
\left[\begin{array}{cc} \mathrm{e}^{-i\frac{g^2t'}{\Delta}} & 0 \\ 0 & 1\end{array}\right],\hspace{1cm}
\left[\begin{array}{cc} \mathrm{e}^{i\frac{g^2t'}{\Delta}} & 0 \\ 0 & 1\end{array}\right],
\end{equation}
respectively.

In \cite{cavity1}, the same method as the one in this section is
suggested for entangled coherent state preparation. They start
with an identical Hamiltonian, but without the restriction on
equal frequencies of the two modes, and perform an adiabatic
elimination, in the original basis, giving the interaction
Hamiltonian
\begin{equation}\label{false}
H_I=\frac{g_1^2}{\Delta_1}a^{\dagger}a\sigma_z+\frac{g_2^2}{\Delta_2}b^{\dagger}b\sigma_z,
\end{equation}
where $\Delta_i=\Omega-\omega_i$, $i=1,2$. The interesting
observation is that the cross terms $a^\dagger b$ and $ab^\dagger$
are omitted in (\ref{false}), with the conclusion that the
elimination has not been carried out correctly. Still, both
Hamiltonians (\ref{effH2}) and (\ref{false}) generate entangled
coherent states, which is a consequence of the relation
(\ref{relation1}). This indicates the importance of this result;
in spite its simplicity, it is nontrivial and a direct consequence
from the properties of coherent states. The correct expression for
the interaction picture Hamiltonian, after the adiabatic
elimination has been performed (analogous to the procedure
presented in the appendix \ref{app}), reads
\begin{equation}\label{correct}
\begin{array}{lll}
H_I & = &
\displaystyle{\frac{g_1^2}{\Delta_1}a^{\dagger}a\sigma_z+\frac{g_2^2}{\Delta_2}b^{\dagger}b\sigma_z}
\\ \\ & & \displaystyle{+\frac{1}{2}\left(\frac{g_1g_2}{\Delta_1}+\frac{g_1g_2}{\Delta_2}\right)\left(a^\dagger
b+ab^\dagger\right)\sigma_z}
\\ \\ & &
\displaystyle{+\frac{1}{2}\left(\frac{g_1^2}{\Delta_1}+\frac{g_2^2}{\Delta_2}\right)\sigma^+\sigma^-,}
\end{array}
\end{equation}
which reproduces the effective Hamiltonian (\ref{effH2}) by
letting $g_2=0$. This is a Hamiltonian for two coupled
oscillators, which may be decoupled by introducing the boson
operators \cite{theoqopt}
\begin{equation}
\tilde{A}=a\cos(\eta)+b\sin(\eta),\hspace{0.6cm}\tilde{B}=-a\sin(\eta)+b\cos(\eta),
\end{equation}
giving
\begin{equation}\label{trans1}
H_I=\left(\lambda\tilde{A}^\dagger\tilde{A}+\zeta\tilde{B}^\dagger\tilde{B}\right)\sigma_z+\frac{1}{2}\left(\lambda+\zeta\right)\sigma^+\sigma^-.
\end{equation}
The parameters are expressed in the old ones as,
\begin{equation}\label{trans2}
\begin{array}{c}
\lambda=\frac{1}{2}\left[\frac{g_1^2}{\Delta_1}\left(1+\cos(2\eta)\right)+\frac{g_2^2}{\Delta_2}\left(1-\cos(2\eta)\right)\right], \\ \\
\zeta=\frac{1}{2}\left[\frac{g_1^2}{\Delta_1}\left(1-\cos(2\eta)\right)+\frac{g_2^2}{\Delta_2}\left(1+\cos(2\eta)\right)\right],
\end{array}
\end{equation}
with
\begin{equation}\label{trans3}
\eta=\frac{1}{2}\tan^{-1}\frac{g_1g_2\Delta_2+g_1g_2\Delta_1}{g_1^2\Delta_2-g_2^2\Delta_1}.
\end{equation}
Thus, the equations (\ref{trans1})-(\ref{trans3}) solves the more
general problem of large, but different, detunings.

\section{Conclusion}\label{conc}
In this paper I have presented a simple model for generation of
highly non-classical entangled  coherent states. A two-level atom
interacts with two cavity modes with equal frequencies, which
under a unitary transformation becomes identical to the standard
JC model describing the interaction between one two-level atom and
one cavity mode. The known analytical results of the original JC
model are used to derive the time-evolution of this extended
system. The special situation of initial coherent states in the
two modes are considered, and it was shown how these may evolve
into entangled coherent states of the modes. Both the zero and
large atom-field detuning situations are studied and the large
amplitude approximations are used in the former. An advantage with
a vanishing detuning is that characteristic interaction times
usually are shorter, giving smaller loses. However, there is a
conflict; for larger amplitudes of the initial coherent states the
approximations are better fulfilled, while the decoherence times
are decreased. The validity of these approximations and
decoherences have been studied in the literature
\cite{gb2,jonasbarry,plk}. The opposite limit of large detuning
was already studied in \cite{cavity1} for generation of entangled
coherent states. They, however, used an effective model that
omitted two essential terms, which will result in new coupling
parameters and new phases of the involved coherent states.

Note that if the amplitudes of the two modes 1 and 2 are equal,
$\alpha=\beta$, and they couple with the same strength to the
atom, $g_1=g_2$, we find $\mu=\sqrt{2}\alpha$ meaning that the
large amplitude approximation works better for quasi mode $I$ than
for the individual modes 1 and 2  separately. The decay rate of
the decoherence terms of the state $\rho$ of a 1-mode
Schr\"odinger cat or a 2-mode entangled coherent state are
proportional to $|\alpha|^2$ or $|\alpha|^2+|\beta|^2$
respectively \cite{qcom,decoh}, where $\alpha$ and $\beta$ are the
amplitudes of the two modes. Thus, they scale the same as for the
situation with $\mu=\sqrt{2}\alpha$ above, indicating that the
model is not more sensitive to decoherence as could be expected
from the two modes included. Schr\"odinger cat states have been
prepared experimentally, both within the large amplitude and the
large detuning approximations, see \cite{gb2,haroche1,haroche2}.
Hence, the experimental verification of entangled coherent states
should be possible within current setups. In addition, the method
presented is as simple as possible; the degrees of freedom is
minimized (no external fields or atomic levels) and as few
experimental steps as possible, which should maximize the
preparation fidelity. However, the measurement process of
entangled coherent states, not discussed in this paper, turns out
to be nontrivial \cite{measure}.

The advantage of the large detuning case is that the amplitudes of
the coherent states, need not be large for the scheme to work. The
model presented in this paper can be generalized to work for
preparing $l$ mode entangled coherent states. Then the
transformation (\ref{trans}) is extended to contain all the $l$
boson operators.

In the model, the two-level atom couples to two cavity modes, so
that the modes must have the same polarizations due to selection
rules. One interesting situation where this is achievable is by
using overlapping cavities \cite{erikajonas}. The atom interacts
with the two modes, belonging to different cavities, in a region
where the cavity fields overlap in space. This has the additional
interesting aspect of entangled states spatially separated. In
\cite{cavity3} a scheme for preparation of entangled coherent
states of separated cavities are presented, by a two-level atom
interacting in succession with the cavities. The advantage with
such a model is that the fields does not need to overlap, while
the disadvantage is that the interaction time increases and
decoherence effects of the atom during the flight between the
cavities may be of importance.


\section*{Acknowledgements}
The author would like to thank Prof. Stig Stenholm for helpful
discussions, and Prof. Peter Knight for mentioning the reference
\cite{plk0}.

\begin{appendix}
\section{Adiabatic elimination}\label{app}
The interaction picture JC Hamiltonian is
\begin{equation}
H=\frac{\Delta}{2}\sigma_z+g\left(a^\dagger\sigma^-+a\sigma^+\right).
\end{equation}
Following \cite{adel1} I introduce the unitary transformation
\begin{equation}
U=\mathrm{e}^S,
\end{equation}
where
\begin{equation}
S=\lambda\left(a\sigma^+-a^\dagger\sigma^-\right),
\end{equation}
for some constant $\lambda$ to be determined later. Using the
operator formula
\begin{equation}
X'=\mathrm{e}^SX\mathrm{e}^{-S}=X+[S,X]+(1/2!)[S,[S,X]]+...\,,
\end{equation}
and keeping terms to second order in the coupling one obtains
\begin{equation}\label{transfor1}
\begin{array}{l}
a'=a+\lambda\sigma^-, \\ \\
\sigma^-=\sigma^-+\lambda a\sigma_z, \\ \\
\sigma_z=\sigma_z-2\lambda\left(a^\dagger\sigma^-+a\sigma^+\right)-2\lambda^2a^\dagger a\sigma_z-\lambda^2\sigma^+\sigma^-.
\end{array}
\end{equation}
By choosing $\lambda=g/\Delta$ the transformed Hamiltonian, to
order $\mathcal{O}(g^2/\Delta)$,  becomes
\begin{equation}
H'=\frac{\Delta}{2}\sigma_z+\frac{g^2}{\Delta}a^\dagger a\sigma_z+\frac{g^2}{2\Delta}\sigma^+\sigma^-.
\end{equation}

\end{appendix}


\pagebreak

\end{document}